\documentclass{emulateapj}

\newcommand{\jcap}{JCAP}

\begin{document}

\title{The Delay of Population~III Star Formation by Supersonic Streaming Velocities}

\author{Thomas~H.~Greif\altaffilmark{1}, Simon~D.~M.~White\altaffilmark{1}, Ralf~S.~Klessen\altaffilmark{2}, Volker~Springel\altaffilmark{3,4}}

\altaffiltext{1}{Max-Planck-Institut f\"{u}r Astrophysik, Karl-Schwarzschild-Stra\ss e 1, 85740 Garching bei M\"{u}nchen, Germany}
\altaffiltext{2}{Zentrum f\"{u}r Astronomie der Universit\"{a}t Heidelberg, Institut f\"{u}r Theoretische Astrophysik, Albert-Ueberle-Stra\ss e 2, 69120 Heidelberg, Germany}
\altaffiltext{3}{Heidelberg Institute for Theoretical Studies, Schloss-Wolfsbrunnenweg 35, 69118 Heidelberg, Germany}
\altaffiltext{4}{Zentrum f\"{u}r Astronomie der Universit\"{a}t Heidelberg, Astronomisches Recheninstitut, M\"{o}nchhofstr. 12-14, 69120 Heidelberg, Germany}

\begin{abstract}
It has recently been demonstrated that coherent relative streaming velocities of order $30\,{\rm km\,s^{-1}}$ between dark matter and gas permeated the universe on scales below a few Mpc directly after recombination. We here use a series of high-resolution moving-mesh calculations to show that these supersonic motions significantly influence the virialization of the gas in minihalos, and delay the formation of the first stars. As the gas streams into minihalos with bulk velocities around $1\,{\rm km\,s^{-1}}$ at $z\sim 20$, the additional momentum and energy input reduces the gas fractions and central densities of the halos, increasing the typical virial mass required for efficient cooling by a factor of three, and delaying Population~III star formation by $\Delta z\simeq 4$. Since the distribution of the magnitude of the streaming velocities is narrowly peaked around a non-negligible value, this effect is important in most regions of the universe. As a consequence, the increased minimum halo mass implies a reduction of the absolute number of minihalos that can be expected to cool and form Population~III stars by up to an order of magnitude. We further find that the streaming velocities increase the turbulent velocity dispersion of the minihalo gas, which could affect its ability to fragment and hence alter the mass function of the first stars.
\end{abstract}


\maketitle

\section{Introduction}
\citet{th10} have recently shown that the acoustic oscillations before recombination give rise to streaming velocities between the dark matter (DM) and gas. Whereas the pressure of the photon-baryon fluid suppressed the growth of baryonic perturbations prior to recombination, the DM aquired streaming velocities of order $v_{\rm stream}\simeq 30\,{\rm km\,s^{-1}}$ at $z=1000$ for typical $1\sigma$ fluctuations.  During recombination, the sound speed of the gas dropped from roughly $c/\sqrt{3}$ to $2\times 10^{-5}c$ as it transformed from a radiation-dominated plasma to a largely neutral gas. As a result, the relative streaming velocities between dark matter and gas transitioned from a subsonic into a supersonic regime, with typical Mach numbers $M\simeq 5$. After this period, the gas remained thermally coupled to the cosmic microwave background (CMB) through Compton scattering with residual free electrons, so that the sound speed decayed as $\propto a^{-1/2}$ until $z\simeq 200$. Disregarding sourcing, the streaming velocities therefore decayed at most as $\propto a^{-1}$, so that the Mach number dropped to $M\ga 2$ once the gas became fully adiabatic, and remained constant thereafter.

Streaming velocities of this magnitude may alter the abundance of DM halos of mass $\sim 10^6\,{\rm M}_{\odot}$ and lead to an increased bias and clustering on scales between a few and $100\,{\rm Mpc}$, roughly corresponding to the Silk damping length and the sound horizon at recombination \citep{th10}. \citet{dps10} argued that the 21 cm signal before and during reionization could also be modified substantially. \citet{mkc11} exploited the coherence of streaming velocities below scales of about $10\,{\rm Mpc}$ to investigate their influence on the formation of the smallest bound objects. They found that the abundance of DM halos with virial masses $10^4\la M_{\rm vir}\la 10^7\,{\rm M}_{\odot}$ was reduced by a few percent, while the gas fractions of these halos were affected by a somewhat larger amount. \citet{sbl11} used smaller box sizes and concentrated on the collapse of gas in minihalos, finding almost no increase in the virial mass and only a mild delay of collapse.

In the present study we also focus on minihalos, where the effect is likely to be most pronounced. Compared to previous investigations, we employ a higher resolution, which allows us to investigate the influence of streaming velocities in more detail, but limits the number of realizations that we can investigate. Furthermore, we here employ the moving mesh code {\small AREPO}, which is based on a set of grid points that are advected with the flow \citep{springel10a}. This allows a higher efficiency and accuracy compared to SPH simulations, while maintaining the natural adaptivity of SPH. Compared to AMR simulations, the Galiliean-invariance of {\small AREPO} is advantageous when dealing with supersonic bulk velocities.

The structure of our work is as follows: In Section~2, we present the numerical setup of the simulations, followed by our main results (Section~3) and a brief discussion of their implications (Section~4). All distances quoted in this paper are in proper units, unless noted otherwise.

\begin{figure*}
\begin{center}
\resizebox{17.4cm}{18.2cm}
{\unitlength1cm
\begin{picture}(17.4,18.2)
\put(0,12.8){\includegraphics[width=5.5cm,height=5.5cm]{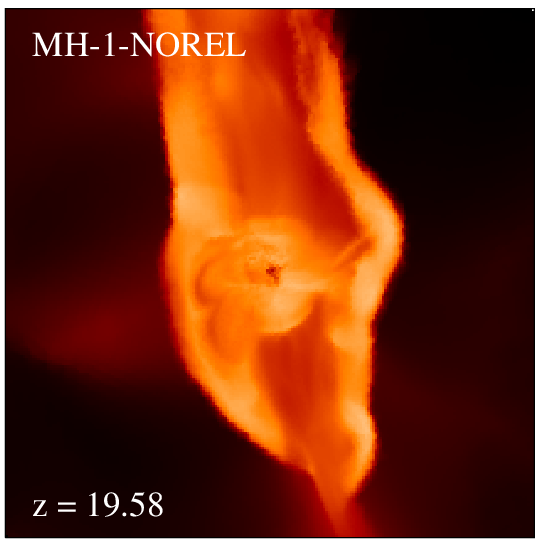}}
\put(5.7,12.8){\includegraphics[width=5.5cm,height=5.5cm]{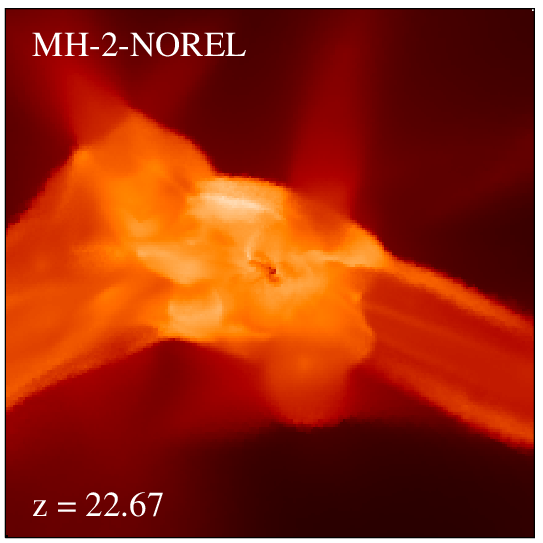}}
\put(11.4,12.8){\includegraphics[width=5.5cm,height=5.5cm]{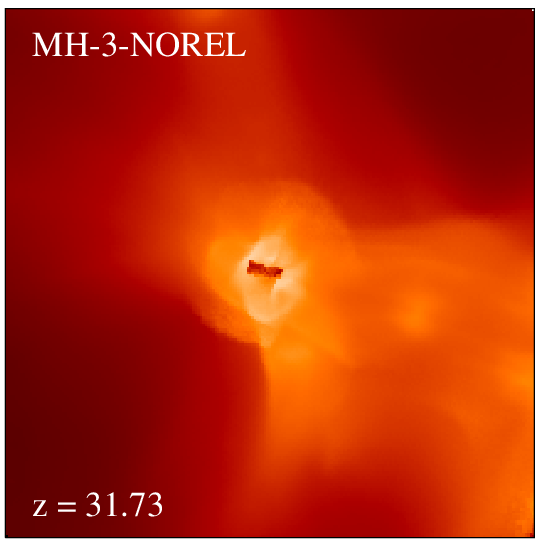}}
\put(0,7.3){\includegraphics[width=5.5cm,height=5.5cm]{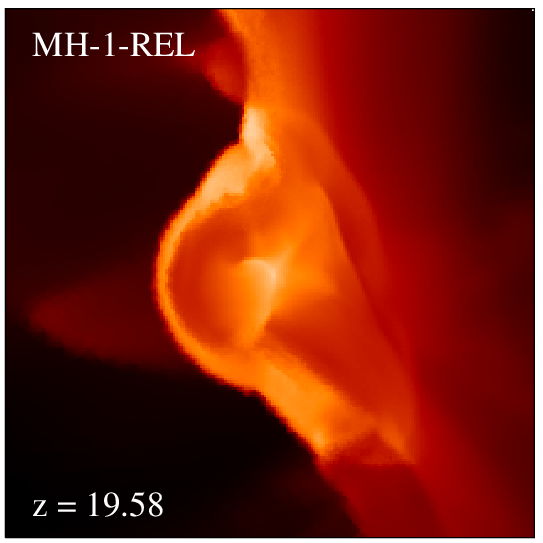}}
\put(5.7,7.3){\includegraphics[width=5.5cm,height=5.5cm]{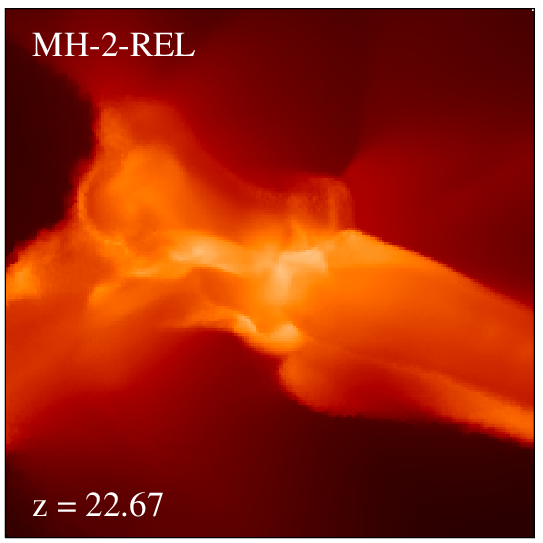}}
\put(11.4,7.3){\includegraphics[width=5.5cm,height=5.5cm]{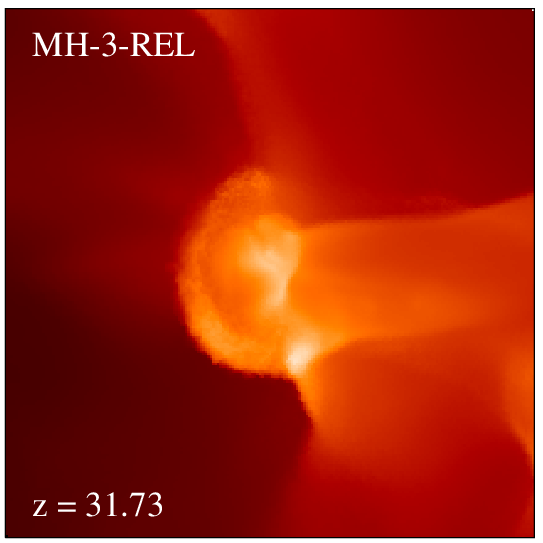}}
\put(0,1.8){\includegraphics[width=5.5cm,height=5.5cm]{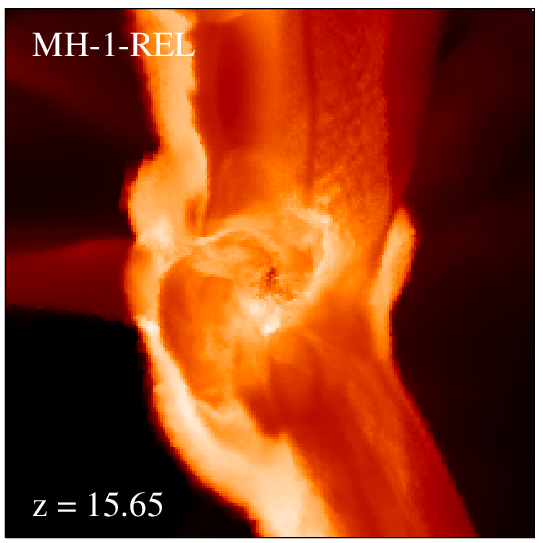}}
\put(5.7,1.8){\includegraphics[width=5.5cm,height=5.5cm]{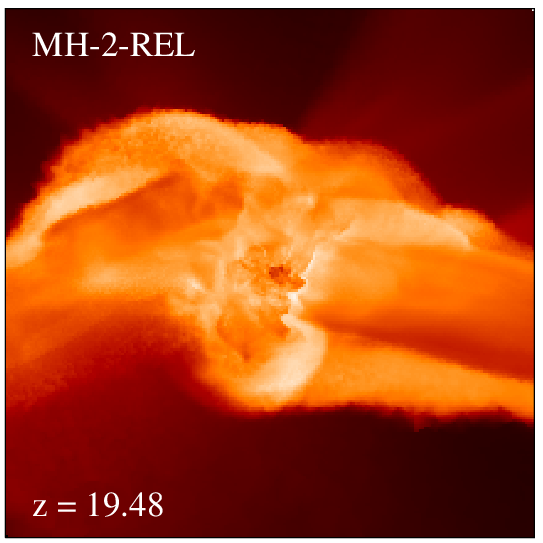}}
\put(11.4,1.8){\includegraphics[width=5.5cm,height=5.5cm]{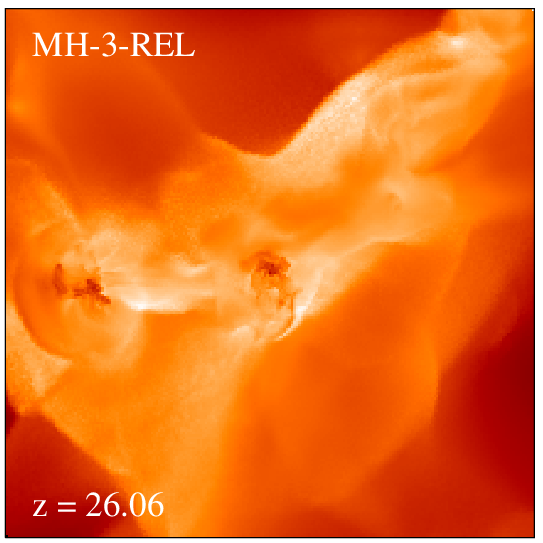}}
\put(0,0){\includegraphics[width=17cm,height=1.5cm]{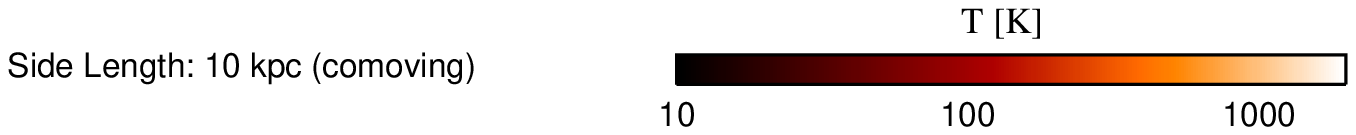}}
\end{picture}}
\caption{A comparison of three statistically independent minihalos with no streaming velocity (top panels), and with an initial streaming velocity of $3\,{\rm km\,s^{-1}}$ applied at $z=99$ from left to right (middle and bottom panels). We show the density-squared weighted gas temperature projected along the line of sight when the hydrogen density in the center has just exceeded $n_{\rm H}=10^9\,{\rm cm}^{-3}$ (top and bottom panels), and when the streaming case has evolved to the same redshift as the no-streaming case (middle panels). In the presence of streaming velocities, the effective Jeans mass of the gas is increased. The underlying DM halo therefore becomes more massive before the gas can cool, which delays the onset of collapse. We also find that virial shocks are more pronounced in the direction of the incoming streaming flow than in other directions. Non-linear effects of this sort may result in a higher velocity disperion of the gas (see also Figure~4).}
\end{center}
\end{figure*}

\begin{figure}
\begin{center}
\includegraphics[width=8cm]{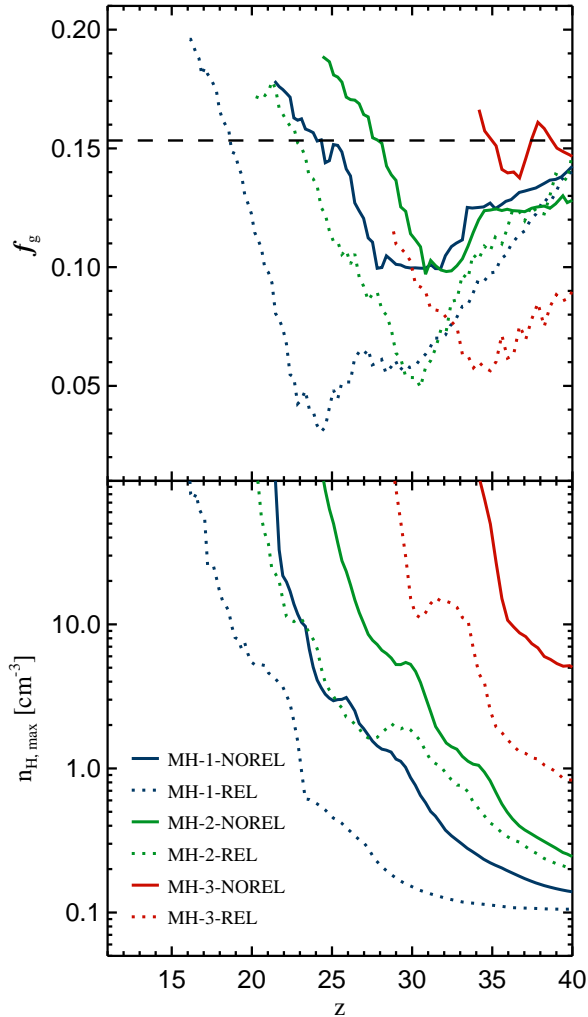}
\caption{The gas fractions and central densities of the minihalos as a function of redshift. The different colors distinguish the individual realizations. The solid and dashed lines denote cases with no streaming velocity and a streaming velocity of $3\,{\rm km\,s^{-1}}$ at $z=99$, respectively. The thick dashed line denotes the cosmological baryon fraction. The increased effective Jeans mass of the gas leads to a reduction of the gas fractions by about $50\%$, which in turn results in a lower central density and a substantial delay of the collapse of the gas.}
\end{center}
\end{figure}

\begin{figure}
\begin{center}
\includegraphics[width=8cm]{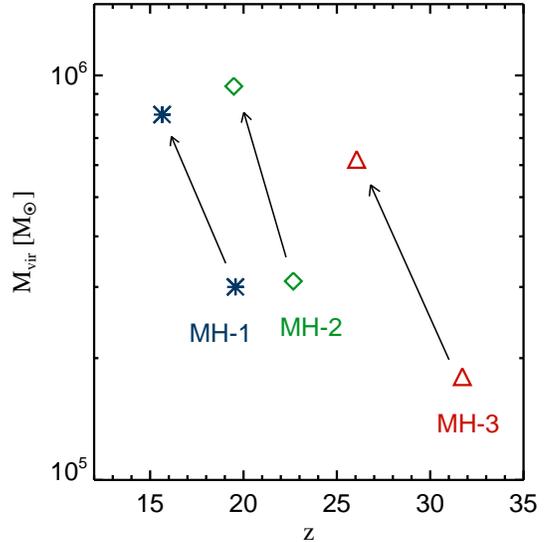}
\caption{The virial masses and collapse redshifts of all minihalos for no streaming velocities (lower symbols), and for an initial streaming velocity of $3\,{\rm km\,s^{-1}}$ at $z=99$ (upper symbols). As indicated by the arrows, the virial mass required for efficient cooling is typically increased by a factor of $\simeq 3$, which delays Pop~III star formation by $\Delta z\simeq 4$.}
\end{center}
\end{figure}

\begin{figure}
\begin{center}
\includegraphics[width=8cm]{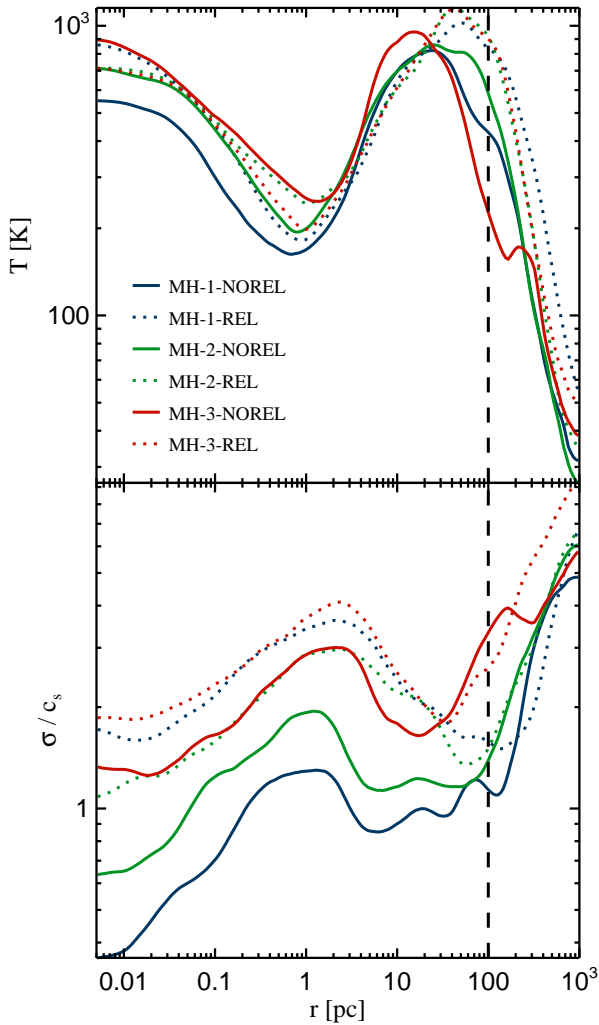}
\caption{The temperature and velocity dispersion of the gas as a function of distance from the center of the halo, shown at a point in time when the hydrogen density has exceeded $n_{\rm H}=10^9\,{\rm cm}^{-3}$. The different colors distinguish the individual realizations, while the solid and dashed lines denote cases with no streaming velocity and a $1\sigma$ streaming velocity of $3\,{\rm km\,s^{-1}}$ at $z=99$, respectively. The black dashed lines denote the approximate virial radius. In cases with a streaming velocity, the increased virial mass leads to a higher virial temperature. This difference is reduced at smaller radii (higher densities), since the gas cools very efficiently during runaway collapse. The bottom panel shows that the velocity dispersion remains systematically higher in cases with a streaming velocity. The increased amount of turbulence might later affect the fragmentation of the gas and alter the mass function of the first stars.}
\end{center}
\end{figure}

\begin{figure}
\begin{center}
\includegraphics[width=8cm]{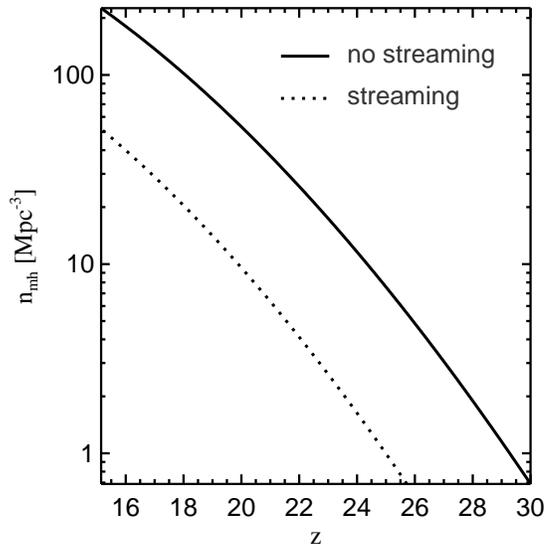}
\caption{The comoving number density of minihalos that are expected to cool and form Pop~III stars for no streaming velocity (solid line), and for an initial streaming velocity of $3\,{\rm km\,s^{-1}}$ at $z=99$ (dotted line). The factor of $\simeq 3$ increase in minimum virial mass leads to a reduction of the number of star-forming minihalos by up to an order of magnitude. The influence of Pop~III stars on observables such as the $21\,{\rm cm}$ background or the reionization of the universe might therefore be substantially reduced.}
\end{center}
\end{figure}

\section{Simulation Setup}

We perform three statistically independent simulations in boxes with a side length of $500\,{\rm kpc}$, denoted MH--1 to MH--3. For each realization of the cosmological density field, we first run a coarse DM simulation to identify the Lagrangian region of the first minihalo that exceeds a virial mass of $5\times 10^5\,{\rm M}_{\odot}$. In these simulations, we employ $256^3$ particles of mass $272\,{\rm M}_{\odot}$ and use a comoving gravitational softening length of $68\,{\rm pc}$. The initial fluctuation power spectrum is that of a $\Lambda$ cold dark matter ($\Lambda$CDM) cosmology at a starting redshift of $z=99$, with matter density $\Omega_m=1-\Omega_\Lambda=0.27$, baryon density $\Omega_b=0.046$, Hubble parameter $h=H_0/100\,{\rm km\,s^{-1}\,Mpc^{-1}}=0.71$ (where $H_0$ is the present Hubble expansion rate), and a spectral index of $n_s=0.96$. These parameters are based on the five-year WMAP results \citep{komatsu09}. The three realizations use different linearly extrapolated present-day normalizations $\sigma_8$, which we set to $\sigma_8=0.9$, $1.1$ and $1.3$, respectively. This is equivalent to altering the redshift of collapse, and allows us to extend the scope of our simulations in light of the limited box sizes employed.

Once the location of a minihalo in each realization has been identified, we use the refinement procedure described in \citet{greif11} to increase the mass resolution within a sufficiently large region around the target halo. The final masses of the DM particles and the mesh-generating points that represent the gas particles are $3.53$ and $0.72\,{\rm M}_{\odot}$, respectively. The comoving gravitational softening length of the refined DM component is $17\,{\rm pc}$. We displace the gas particles relative to the DM particles by half the mean interparticle separation, implicitly assuming that the transfer functions of the gas and DM components are equivalent. This can substantially affect the filtering mass scale \citep{nyb10}, but we chose to neglect this caveat since we here investigate the differences between simulations with and without streaming velocities, but otherwise identical parameters.

In simulations MH--1--REL to MH--3--REL, the gas particles obtain an additional velocity of $v_{\rm stream}=3\,{\rm km\,s^{-1}}$ in the $x$-direction, which approximately corresponds to a $1\sigma$ peak. The high resolution region is extended in the $x$-direction to ensure that only refined particles collapse into the minihalos. Simulations without an additional streaming velocity are denoted MH--1--NOREL to MH--3--NOREL. The chemical and thermal evolution of the gas is captured with a primordial chemistry network that tracks the abundances and the cooling from H, H$^+$, H$^-$, H$_2^+$, H$_2$, He, He$^+$, He$^{++}$, D, D$^+$, HD, and free electrons \citep{gj07,greif11}. We account for all relevant cooling processes, including H$_2$ and HD rotational and vibrational line cooling, H$_2$ collision-induced emission (CIE) cooling, H$_2$ collisional dissociation cooling, and heating due to three-body H$_2$ formation. For further details see \citet{greif11}.

\section{Results}

\subsection{Delay of Runaway Collapse}

To first order, streaming velocities may be considered as an additional source of pressure, which increases the Jeans mass of the gas:
\begin{equation}
M_{\rm J}=\frac{v_{\rm eff}^3}{2\,G^{3/2}\rho^{1/2}}\mbox{\ ,}
\end{equation}
with $v_{\rm eff}^2=c_{\rm s}^2+v_{\rm stream}^2$. According to this simple argument, the minimum halo mass required for collapse is increased. \citet{sbl11} showed that this picture qualitatively describes their simulation results. The effect is particularly important in minihalos, since a $1\sigma$ streaming velocity of $1\,{\rm km\,s^{-1}}$ at $z\simeq 20$ is comparable to the sound speed of virialized gas in halos with $M_{\rm vir}\simeq 1.5\times 10^5\,{\rm M}_{\odot}$ -- a lower limit for Pop~III star formation \citep{mba01,on07}.

In Figure~1, we show the temperature of the gas once the central hydrogen density first exceeds $n_{\rm H}=10^9\,{\rm cm}^{-3}$. We define this as the collapse redshift, since at this point in time the central gas cloud has evolved well beyond the initial Jeans instability and the density increases exponentially. In cases with a streaming velocity, the minihalos are more massive at the point of collapse, implying a higher virial temperature. The motions accompanying their virialization also appear to be more complex. For example, the virial shocks in the direction of the incoming streaming flows are more pronounced than on the opposite side. This effect is particularly evident in simulation MH--1, where a filament was by chance aligned perpendicular to the streaming velocity.

In Figure~2, we show the gas fractions and maximum densities of the three minihalos as a function of time. In cases with an imposed streaming velocity, the gas fractions are reduced by about $50\%$. This agrees well with the analytical predictions of \citet{tbh10}. At a given point in time, the maximum density of the halo is reduced, so that the halo does not cool until it has accreted enough mass to compensate for this effect. The resulting shifts in virial mass and redshift are shown in Figure~3. The virial masses increase from $3.0$, $3.1$, and $1.8\times 10^5\,{\rm M}_{\odot}$ to $8.0$, $9.4$, and $6.2\times 10^5\,{\rm M}_{\odot}$, respectively, corresponding to factors of $2.7$, $3.0$, and $3.4$. The redshifts of collapse are shifted from $19.6$, $22.7$, and $31.7$ to $15.6$, $19.5$, and $26.1$, resulting in delays of $\Delta z=4$, $3.2$, and $5.5$. We note that the factor of $\simeq 3$ increase in virial mass is particularly important since it non-linearly affects the number density of DM halos that can cool and form Pop~III stars (see Section~3.3).

\subsection{Thermal Evolution and Turbulence}

The increased momentum and energy input by streaming velocities also affects the thermal and turbulent evolution of the gas. In Figure~4, we compare the temperature and velocity dispersion within all three minihalos. In cases with a streaming velocity, the gas temperature near the virial radius is systematically higher. This is not suprising, since the virial temperature scales with virial mass and redshift as \begin{equation}
T_{\rm vir}\simeq 10^3\,{\rm K}\left(\frac{M_{\rm vir}}{10^6\,{\rm M}_{\odot}}\right)^{2/3}\left(\frac{1+z}{15}\right)\mbox{\ .}
\end{equation}
A factor of $\simeq 3$ increase in the virial mass and a comparatively small decrease in redshift therefore result in a factor of $\sim 2$ higher virial temperature. Once the gas begins to cool efficiently, this difference is quickly reduced, however, and we find no clear difference remaining below $r\simeq 10\,{\rm pc}$.

In the bottom panel of Figure~4, we also show the velocity dispersion of the gas in radial bins after subtracting the bulk velocity of the halo and a spherically averaged radial infall and rotation velocity. Unlike the temperature, the velocity dispersion remains systematically higher by a factor of a few even at very high densities. Evidently, the gas becomes more turbulent in the presence of streaming velocities. This could affect the mass function of the first stars, since a higher ratio of the velocity dispersion to the sound speed usually results in increased fragmentation at later stages in the collapse \citep{clark11}. As a result, the typical mass of a Pop~III star may be reduced.

\subsection{Abundance of Minihalos}

How common is the above delay considering the minihalo population as a whole? Since each component of the large-scale streaming velocity is distributed according to a Gaussian, the magnitude of the velocity vector obeys a Maxwell distribution \citep{tbh10}:
\begin{equation}
p_{\rm stream}=\left(2\pi\sigma_{1{\rm d}}^2\right)^{-3/2}4\pi v_{\rm stream}^2\exp{\left(-\frac{v_{\rm stream}^2}{2\sigma_{1{\rm d}}^2}\right)}\mbox{\ ,}
\end{equation}
where $\sigma_{1{\rm d}}$ is the dispersion of each individual velocity component. The fraction of the universe with streaming velocities greater than $1.5\,{\rm km\,s^{-1}}$ at $z=99$, which we consider a lower limit for the above delay to be significant, may be found by integrating the above function from $\sigma/2=\sigma_{1{\rm d}}\sqrt{3}/2$ to infinity, which yields approximately $0.86$. This shows that our results may be considered representative for most of the volume.

The cosmological number density of minihalos hosting Pop~III stars may then be estimated using the Sheth-Tormen \citep{smt01} mass function:
\begin{equation}
n_{\rm mh}(z)=\int_{M_{\rm min}}^{M_{\rm max}}n_{\rm st}(M,z)\,dM\mbox{\ ,}
\end{equation}
where we set $M_{\rm min}=1.5\times 10^5\,{\rm M}_{\odot}$ for the case of no streaming velocities and $M_{\rm min}=5\times 10^5\,{\rm M}_{\odot}$ for the case of a universal $1\sigma$ streaming velocity, representing the factor of $\simeq 3$ increase in minimum halo mass. The resulting number densities should be considered upper limits, since not every halo at the low-mass end forms a Pop~III star. We set $M_{\rm max}=10^8\,{\rm M}_{\odot}$, but note that our results are not sensititive to this parameter, since massive halos are rare. As shown in Figure~5, the number of minihalos that cool and form stars is reduced by up to an order of magnitude in the presence of streaming velocities. Such a large effect implies that streaming velocities should be taken into account when the influence of the first stars on observables is investigated.

\section{Discussion}

We have found that supersonic streaming velocities between the DM and gas substantially delay the onset of gravitational collapse in minihalos. The virial mass required for efficient cooling is increased by a factor of $\simeq 3$, which results in an average delay of Pop~III star formation by $\Delta z=4$. Streaming velocities also enhance the build-up of turbulence during runaway collapse, which could affect the fragmentation of the gas and hence the mass function of the first stars.

Our results agree well with the simulations presented in \citet{sbl11} concerning the delay of collapse. In addition, and in contrast to their result, we find a significant increase in the minimum halo mass required for the gas to cool. This might be due to the $\simeq 10$ times higher resolution employed in the present study, ensuring that the DM softening length is comparable to the size of the Jeans-unstable gas cloud. If this criterion is not fulfilled, the collapse of the gas might be artificially delayed \citep[e.g.,][]{yoshida03a}. We have investigated this for the case MH--1. We increased the initial DM and gas particle masses by factors of $8$ and $64$ to $28.2$ and $225\,{\rm M}_{\odot}$ for the DM and $5.8$ and $46.4\,{\rm M}_{\odot}$ for the gas. In these simulations we also decreased the number of cells enforced per Jeans length from $128$ to $64$ and $32$ \citep[for details, see][]{greif11}. When decreasing the resolution, the virial mass of the minihalo at the point of collapse is increased from $3.0$ to $3.3$ and $4.7\times 10^5\,{\rm M}_{\odot}$, and the redshift of collapse is shifted from $19.6$ to $19.2$ and $18.1$. On the other hand, an increase of the initial resolution by a factor of eight results only in a marginal decrease of the virial mass to $2.9\times 10^5\,{\rm M}_{\odot}$, and shifts the redshift of collapse to $19.8$. This shows that insufficient resolution can artificially delay the collapse of the halo, which might be one of the reasons why \citet{sbl11} find significantly higher halo masses.

Due to the large shift in mass in the presence of streaming velocities, the number density of minihalos expected to form Pop~III stars is reduced by up to an order of magnitude, and is further modulated with a spatial coherence length corresponding to many Mpc in the present universe. This could leave a distinct imprint on the 21 cm background and the reionization of the universe \citep{furlanetto06,bl01}, as well as affect the observability of Pop~III supernovae \citep[e.g.][]{bl02}. Ultimately, this modulation might be inherited by the distribution of more massive objects forming at later times, thereby biasing large-scale structure estimates of cosmological parameters such as those related to the properties of dark energy.

\acknowledgements{T.H.G would like to thank Tom Abel, Naoki Yoshida, Paul Clark and Simon Glover for many stimulating discussions. R.S.K. acknowledges financial support from the Baden-W\"{u}rttemberg Stiftung via their program International Collaboration II (grant P-LS-SPII/18) and from the German Bundesministerium f\"{u}r Bildung und Forschung via the ASTRONET project STAR FORMAT (grant 05A09VHA). We thank the Leibniz Supercomputing Center of the Bavarian Academy of Sciences and Humanities, and the Computing Center of the Max-Planck-Society in Garching, where the simulations were carried out.}

\end{document}